\documentclass[preprint,aps,12pt,showpacs,nofootinbib,tightenlines]{revtex4}
\usepackage{amsmath}
\usepackage{amssymb}
\usepackage{epsfig}
\usepackage{graphicx}
\begin{document}
\newcommand{\be}{\begin{equation}}
\newcommand{\ee}{\end{equation}}
\newcommand{\bea}{\begin{eqnarray}}
\newcommand{\eea}{\end{eqnarray}}
\newcommand{\nnb}{\nonumber}
\def \epjc{  Eur. Phys. J. C }
\def \jpg{  J. Phys. G }
\def \jhep{  J. High Energy Phys. }
\def \npb{  Nucl. Phys. B }
\def \plb{  Phys. Lett. B }
\def \prd{  Phys. Rev. D }
\def \prl{  Phys. Rev. Lett.  }
\def \rmp{  Rev. Mod. Phys. }

\title{$ B^0_{d,s} - \bar{B}^0_{d,s} $ mixing in the model III 2HDM}
\author{Chao-Shang Huang$^a$}
\email{csh@itp.ac.cn}
\author{Jian-Tao Li$^{a,b}$}
\email{ljiantao@itp.ac.cn} \affiliation{$^a$ Institute of
Theoretical Physics, Academia Sinica, Beijing 100080, China\\
$^b$ Graduate School of the Chinese Academy of Science, YuQuan
Road 19A, Beijing 100039, China}

\date{\today}

\begin{abstract}
In the model III 2HDM there are new CP violating phases which
would affect the $B_{d,s}^0 - \bar{B}_{d,s}^0$ mixing. In this
paper, we calculate the new physics contributions to the neutral B
meson mass splitting $\Delta M_{B_q}$ (q=d, s) at the
next-to-leading order (NLO) level. Using the high accuracy data
and other relevant data, we draw the constraints on the parameter
space of the model III 2HDM. Moreover, we calculate the new
physics corrections to the ratio $q/p$. It is found that the phase
of $(q/p)_n$ for $B_d$ which is due to the new contributions is
very small and consequently in agreement with the measurements of
the time dependent CP asymmetry $S_{J/\psi K}$ in $B\to J/\psi
K_S$. On the contrary, the phase of $(q/p)_n$ for $B_s$ is large
enough to give significant effects on CP violation in the neutral
$B_s$ system.

\end{abstract}

\pacs{12.15.F, 12.60.F, 12.60.J, 11.30.E}

\maketitle


\newpage
\section{introduction}\label{sec-1}

The neutral B-meson system has been of fundamental importance in
testing the Standard Model (SM) picture of flavor changing neutral
currents (FCNC) and CP violation and in probing virtual effects
from potential new physics beyond SM at low energies. The
presentence of CP violation in the neutral B meson system has been
established. The measured value~\cite{babar,belle,cdf} of the time dependent CP
asymmetry $S_{J/\psi K}$ in $B\to J/\psi K_S$
\begin{equation}
S_{J/\psi K}=\sin (2 \beta (J/\psi K_{S}))_{\rm world-ave}=0.734 \pm 0.054.
\label{sjp}
\end{equation}
is in agreement with the prediction in the standard model (SM).

In the $B_{\sl q}^0$ ({\sl q}=d,\,s) system, we expect model
independently that~\cite{cpviolation}
\begin{equation}
\Gamma_{12} \ll M_{12}
\end{equation}
where $\Gamma_{12}$ and $M_{12}$ are the off-diagonal terms of $2
\times 2$ decay width matrix and mass matrix of $B_{\sl
q}^0-\bar{B}_{\sl q}^0$. Thus one has
\begin{eqnarray}
\frac{q}{p}&=&-\frac{M_{12}^\ast}{|M_{12}|}, \\
\Delta M_{B_{\sl q}}&=&2 |M_{12}|,
\label{msd}
\end{eqnarray}
where $q$ and
$p$ are defined as
\begin{equation}
|B_{L,H}\rangle = p|B^0\rangle \pm q|\bar{B}^0\rangle
\end{equation}
with $B_{L,H}$ denoting the light and heavy meson mass
eigenstates, and the normalization condition is
\begin{equation}
|p|^2+|q|^2=1.
\end{equation}

In the SM, one has, according to the box diagram calculation,
\begin{equation}\label{qpd}
\frac{q}{p}=-\frac{V_{tb}^\ast V_{t{\sl q}}}{V_{tb}V_{t{\sl
q}}^\ast}.
\end{equation}
and the deviation of $|q/p|$ from 1 is $\sim 10^{-3}$ ($10^{-5}$)
for $B_d$ ($B_s$)~\cite{nir} which is unobservably small. Eq.
(\ref{qpd}) for the $B_d^0$ system has been verified by
measurements of the time dependent CP asymmetry $S_{J/\psi K}$ in
$B\to J/\psi K_S$, as mentioned above. Therefore, it would give a
constraint on parameters of new theoretical models.

As it is obvious from Eq. (\ref{msd}), $B^0 - \bar{B}^0$ mixing is
responsible for the small mass differences between the mass
eigenstates of neutral B mesons. The results of measurements of
the mass splitting $\Delta m_d$ have been obtained with high
accuracy. The current world averages of $\Delta m_q$ ($q=d,\,s$)
are as follow \cite{pdg2003,hfag03}
\begin{eqnarray}
\Delta M_{B_d} &=& 0.502 \pm 0.007 \, ps^{-1} \nonumber\\
\Delta M_{B_s} &>& 14.4\, ps^{-1}
\end{eqnarray}
which would impose a stringent constraint on new model building.

In the SM, $B^0 - \bar{B}^0$ mixing is dominated by the box
diagrams with two internal t-quarks and W gauge bosons. In new
physics models, the box diagrams with one or two W gauge bosons
replaced by the new charged scalars or vector bosons and/or top
quarks replaced by new fermions can also contribute to $B^0 -
\bar{B}^0$ mixing. Using the precision data we can put stringent
constrains on the parameter space of a new physics model. In the
paper we study $B^0 - \bar{B}^0$ mixing in the model III 2HDM.

It is well-known that in the model III 2HDM the couplings
involving Higgs bosons and fermions can have complex phases, the
new CP violating phases would affect $B^0 - \bar{B}^0$ mixing.
$B^0 - \bar{B}^0$ mixing has extensively been studied in the model
III 2HDM during the past years. The charged-Higgs boson
contributions to $B^0 - \bar{B}^0$ mixing have been calculated at
the leading order for a long time~\cite{asw80}. In the framework
of SM, Ref.~\cite{buras90} is the first one to present the NLO QCD
corrections to $B^0 - \bar{B}^0$ mixing while in conventional
model I and model II 2HDM Ref.~\cite{urban98} is. The possible
constraints on the model III 2HDM from the measured parameter $x_d
= \Delta M_B /\Gamma_B$ were studied, for example, in
Refs.~\cite{grant95,atwood97,bck} at the LO level. At the NLO
level, Ref.~\cite{xiao} has calculated the charged Higgs boson
contribution to mass splitting $\Delta M_{B_d}$ and drawn the
constraints on the parameters of model III in terms of the high
precision data, considering the uncertainty of the
non-perturbative parameter $f_{B_d}\sqrt{\hat{B}_{B_d}}$. The
effects of the new CP violating phases on $\Delta M_{B_d}$ have
been taken into account \cite{bck}. However, in Ref.~\cite{bck} as
well as Ref.~\cite{xiao} all new parameters except for
$\lambda_{tt,bb}$ in the model III 2HDM are set to be zero.
Furthermore, the ratio $q/p$ which is important ingredient to
study CP violation in the neutral B system has not been analyzed.
In this paper, we will calculate the new physics contributions to
$\Delta M_{B_q}$ taking into account the effects of the complex
phases and keeping all relevant parameters of the model, which are
only subjective to constraints from experiments, non zero at the
NLO level in the model III 2HDM. By comparing the theoretical
predictions with the high accuracy data, we draw the constraints
on the free parameters of model III. Moreover, by using the
constrained parameters we calculate the contributions of new
physics to the ratio $q/p$ in the neutral B system.

The organization of this paper is as follows. In the next section,
we describe the model III 2HDM briefly. In Section III, we give
the effective Hamiltonian responsible for $B^0 - \bar{B}^0$ mixing
and calculate the mass splitting $\Delta M_{B_q}$ and the ratio
$q/p$ at the NLO level in the model III. The section IV is devoted
to numerical results. Finally, we conclude in section V.


\section{The model III two higgs doublet model}\label{sec-2}

As the simplest extension of the SM, the so-called
two-Higgs-doublet models \cite{2hdm,gunion90} may naturally have
flavor changing neutral currents (FCNC's) mediated by the Higgs
bosons at the tree-level, unless an {\it ad hoc} discrete symmetry
is imposed. In this paper, we shall focus on the model III 2HDM
\cite{sher87,hou92,atwood97}. In the model III, there is no
discrete symmetry and both the Higgs doublets can couple to the
up-type and down-type quarks. In general one can have a Yukawa
Lagrangian of the form
\begin{equation}
{\cal L}_{Y}= \eta^{U}_{ij} \bar Q_{i,L} \tilde H_1 U_{j,R} +
\eta^D_{ij} \bar Q_{i,L} H_1 D_{j,R} + \xi^{U}_{ij} \bar
Q_{i,L}\tilde H_2 U_{j,R} +\xi^D_{ij}\bar Q_{i,L} H_2 D_{j,R}
\,+\, h.c. \label{lyukmod3}
\end{equation}
where $H_i$ ($i=1,2$) are the two Higgs doublets, while $\eta^{U,D}_{i,j}$
and $\xi^{U,D}_{i,j}$ ($i,j=1,2,3$ are family index) are the nondiagonal
matrices of the Yukawa couplings. We can choose to express $H_1$ and $H_2$
in a suitable basis such that only the $\eta_{ij}^{U,D}$ couplings
generate the fermion masses, i.e.
\begin{equation}
 \langle H_1\rangle=\left(\begin{array}{c} 0\\
 {v\over \sqrt{2} } \end{array}\right)
\,\, , \,\,\,\,\,\,\,\, \langle H_2\rangle=0
\end{equation}
Then the two Higgs doublets in the basis are of the form
\begin{equation}
 \label{base} H_1=\frac{1}{\sqrt{2}}\left[\left(\begin{array}{c} 0 \\
v+\phi^0_1 \end{array}\right)+ \left(\begin{array}{c} \sqrt{2}\, G^+\\
i G^0\end{array}\right)\right]\,\,, \,\,\,\,\,\,\,\,
H_2=\frac{1}{\sqrt{2}}\left(\begin{array}{c}\sqrt{2}\,H^+\\ \phi^0_2+i
A^0\end{array}\right)
\end{equation}
where $H^\pm$ are the physical charged-Higgs bosons and $A^0$ is
the physical CP-odd neutral Higgs boson, $G^0$ and $G^\pm$ are the
Goldstone bosons that would be eaten away in the Higgs mechanism
to become the longitudinal components of the weak gauge bosons.
The advantage of using the basis is that the first doublet $H_1$
corresponds to the scalar doublet of the SM while the new Higgs
fields arise from the second doublet $H_2$. The $\phi^0_1$ and
$\phi^0_2$ are not the neutral mass eigenstates but linear
combinations of the CP-even neutral Higgs boson mass eigenstates,
$H^0$ and $h^0$:
\begin{eqnarray}
\label{masseigen}
H^0 & = & \phi^0_1 \cos\alpha + \phi^0_2\sin\alpha \\
\nonumber h^0 & = & -\phi^0_1\sin\alpha + \phi^0_2 \cos\alpha
\end{eqnarray}
 where $\alpha$ is the mixing angle, such that for $\alpha\!=\!0$,
($\phi^0_1$, $\phi^0_2$) coincide with the mass eigenstates.

Note that $Q_{i,L}$, $U_{j,R}$ and $D_{j,R}$ in Eq.(\ref{lyukmod3})
are weak eigenstates, which can be rotated into mass eigenstates.
After the transformation the flavor changing (FC) part of the Yukawa
Lagarangian becomes \cite{hll}
\begin{equation}
{\cal L}_{Y,FC} = \hat\xi^{U}_{ij} \bar Q_{i,L}\tilde H_2
U_{j,R}
 +\hat\xi^D_{ij}\bar Q_{i,L} H_2 D_{j,R} \,+\, h.c. \label{lyukfc}
\end{equation}
where
\begin{equation}
\hat\xi^{U,D}=(V_L^{U,D})^{-1}\cdot \xi^{U,D} \cdot V_R^{U,D}
\label{neutral}
\end{equation}
in which $V_{L,R}^{U,D}$ are the rotation matrices acting on the
up- and down-type quarks, with left or right chirality
respectively, so that $V_{CKM}=(V_L^U)^{\dag}V_L^D$ is the usual
Cabibbo-Kobayashi-Maskawa (CKM) matrix. Feynman rules of Yukawa
couplings follows from Eq. (\ref{lyukfc}) can be found in, e.g.,
Refs. \cite{atwood97,bck}. The FCNC couplings are given by the
matrices  $\hat\xi^{U,D}$ and the charged FC couplings are given
by
\begin{eqnarray}
\hat\xi^{U}_{\rm charged}\!&=&\!\hat\xi^{U}\cdot
V_{CKM}\nonumber\\
\hat\xi^{D}_{\rm charged}\!&=&\!V_{CKM}\cdot \hat\xi^{D}
\label{charged}
\end{eqnarray}
Because the definition of the $\xi^{U,D}_{ij}$ couplings is arbitrary,
we can take the rotated couplings as the original ones and shall write
$\xi^{U,D}$ in stead of $\hat{\xi}^{U,D}$ hereafter. The Cheng-Sher
ansatz for $\xi^{U,D}$ is \cite{sher87}
\begin{equation}
\xi^{U,D}_{ij}=\frac{ g\,\sqrt{m_im_j}}{\sqrt{2}\,M_W } \lambda_{ij},
\label{lij}
\end{equation}
by which the quark-mass hierarchy ensures that the FCNC within the
first two generations are naturally suppressed by the small quark
masses, while a larger freedom is allowed for the FCNC involving
the third generation. In the ansatz the residual degree of
arbitrariness of the FC couplings is expressed through the
$\lambda_{ij}$ parameters which are of order one and need to be
determined by the available experiments, and they may also be
complex. In the paper we choose $\xi^{U,D}$ to be diagonal and set
the masses of u and d quark to be zero for the sake of simplicity,
so that $\lambda_{ii}$ ($i=c,s,t,b$) are the new free
parameters and will enter into the Wilson coefficients relevant to
the process.

\section{$B^0 - \bar{B}^0$ mixing in the model III 2HDM}\label{sec-3}

\subsection{Effective Hamiltonian}

The most general effective Hamiltonian for $\Delta B = 2$ processes beyond
the SM can be written as
\begin{eqnarray}
{\cal H}_{\rm eff}^{\Delta B=2} &=& \sum_{i=1}^{5} C_i Q_i +
  \sum_{i=1}^{3} \tilde{C}_i \tilde{Q}_i
\end{eqnarray}
with
\begin{eqnarray}
Q_1 &=& \bar{q}^{\alpha}_L \gamma^\mu b^{\alpha}_L \bar{q}^{\beta}_{L}
  \gamma_\mu b^{\beta}_L \nonumber\\
Q_2 &=& \bar{q}^{\alpha}_R  b^{\alpha}_L \bar{q}^{\beta}_R b^{\beta}_L
\nonumber\\
Q_3 &=& \bar{q}^{\alpha}_R  b^{\beta}_L \bar{q}^{\beta}_R b^{\alpha}_L
\nonumber\\
Q_4 &=& \bar{q}^{\alpha}_R  b^{\alpha}_L \bar{q}^{\beta}_L b^{\beta}_R
\nonumber\\
Q_5 &=& \bar{q}^{\alpha}_R  b^{\beta}_L \bar{q}^{\beta}_L b^{\alpha}_R
\end{eqnarray}
where $q = s,\,d$, corresponding to the operators of $B_s$ and
$B_d$ system respectively, $P_{L,R} \equiv (1 \mp \gamma_5)/2$,
and $\alpha$, $\beta$ are colour indexes. The tilde operators
$\tilde{Q}_i$ ($i=1,2,3$) are obtained from $Q_i$ correspondingly
by exchanging $L \leftrightarrow R$.

At one loop level only the charged scalars are relevant for the
box diagrams contributing to the $\bar{B}^0 - B^0$ mixing
amplitude. Using the Cheng-Sher ansatz for $\xi^{U,D}$ which we
assume in the paper, we rewrite the tree level couplings of $H_l^+
\equiv ( H^+, G^+)$~\cite{notation} (where $G^\pm$ is the would-be
Goldstone boson) as follows
\begin{equation}
{\cal{L}}_{int} = H^+_l \bar{u}_A V_{AI} (Y_L^{AIl} P_L + Y_R^{AIl} P_R )
d_I + h.c.
\end{equation}
where
\begin{displaymath}
Y_L^{AIl} = \frac{e}{\sqrt{2} s_W} \frac{m_{u_A}}{M_W} \times
\left\{ \begin{array}{ll}
\,\,\lambda_{AA}^\ast \hspace{1cm} & \textrm{ for \, l=1} \\
\,\,\,\,\, 1 \hspace{2cm} & \textrm{ for \, l=2}
\end{array} \right.
\end{displaymath}

\begin{displaymath}
Y_R^{AIl} = \frac{e}{\sqrt{2} s_W} \frac{m_{d_I}}{M_W} \times
\left\{ \begin{array}{ll}
\,\,-\lambda_{II} \hspace{1cm} & \textrm{ for \, l=1} \\
\,\,\,\,\, -1 \hspace{2cm} & \textrm{ for \, l=2}
\end{array} \right.
\end{displaymath}
with $A=t,c,u$ and $I=b,s,d$.

In terms of the coefficients $Y_L^{AIl} $ and $Y_R^{AIl} $ we can
express the contributions of $H_l^+$ to the Wilson coefficients
$C_i$ of the relevant operators responsible for $B^0 - \bar{B}^0$
mixing as follows.

Diagrams with one $W^\pm$ and one $H^\pm$ give\footnote{The contribution of
$G^\pm$ to $C_1(\mu)$ is already taken into account in the Inami-Lim function
$S_0(x_t)$~\cite{Inami-Lim}. Masses of the $u$ and $c$ quarks are neglected.}:
\small
\begin{eqnarray}
&&C_1(\mu)=\frac{(V_{AI}^\ast V_{AJ})^2}{16 \pi^2}{e^2\over2s^2_W}
\sum_A m_{u_A} ^2 Y_L^{\dagger AI1} Y_L^{AJ1}
D_0(m_{u_A}^2 ,m_{u_A}^2,M_W^2,M_{H^+}^2 )\phantom{aa}\nonumber\\
&&C_2(\mu)=\frac{(V_{AI}^\ast V_{AJ})^2}{16 \pi^2}
{e^2\over2s^2_W}\sum_A\sum_{l=1}^2 Y_R^{\dagger AIl} Y_R^{AJl}4
D_{00}(m_{u_A}^2,m_{u_A}^2,M_W^2,m_{H_l^+}^2)
\end{eqnarray}
\normalsize where $s_W \equiv \sin \theta_W$ ($\theta_W$ is the
Weinberg angle), the definition of the four-point integral
functions $D_0$ and $D_{00}$ can be found in the Appendix.

Diagrams with two $H_l^\pm$ give\footnote{In the sum over $l$ and
$n$ in the expression for $C_1(\mu)$ the contribution of $G^\pm
G^\mp$ is excluded since it has been taken into account in the
function $S_0(x_t)$~\cite{Inami-Lim}.}: \small
\begin{eqnarray}
\label{eq:wilson}
&&C_1(\mu) = -{1\over2}\frac{(V_{AI}^\ast V_{AJ})^2}{16 \pi^2}\sum_A\sum_{l,n}
Y_L^{\dagger AIl} Y_L^{AJn} Y_L^{\dagger AIn} Y_L^{AJl}
D_{00}(m_{u_A}^2,m_{u_A}^2,M_{H^+_l}^2,M_{H^+_n}^2)\phantom{aa}\nonumber\\
&&\tilde{C}_1(\mu) = -{1\over2}\frac{(V_{AI}^\ast V_{AJ})^2}{16 \pi^2}\sum_A
\sum_{l,n}^2 Y_R^{\dagger AIl} Y_R^{AJn} Y_R^{\dagger AIn} Y_R^{AJl}
D_{00}(m_{u_A}^2,m_{u_A}^2,M_{H^+_l}^2,M_{H^+_n}^2)
\phantom{aa}\nonumber\\
&&C_2(\mu) = -{1\over2}\frac{(V_{AI}^\ast V_{AJ})^2}{16 \pi^2}\sum_A
\sum_{l,n}^2 m_{u_A}^2 Y_R^{\dagger AIl} Y_L^{AJn} Y_R^{\dagger AIn} Y_L^{AJl}
D_0(m_{u_A}^2, m_{u_A}^2, M_{H^+_l}^2, M_{H^+_n}^2)\phantom{aa}\nonumber\\
&&\tilde{C}_2(\mu) = -{1\over2}\frac{(V_{AI}^\ast V_{AJ})^2}{16 \pi^2}\sum_A
\sum_{l,n}^2 m_{u_A}^2 Y_L^{\dagger AIl} Y_R^{AJn} Y_L^{\dagger AIn} Y_R^{AJl}
D_0(m_{u_A}^2, m_{u_A}^2, M_{H^+_l}^2, M_{H^+_n}^2)\phantom{aa}\nonumber\\
&&C_4(\mu) =-\frac{(V_{AI}^\ast V_{AJ})^2}{16 \pi^2}\sum_A\sum_{l,n}^2
m_{u_A}^2 Y_L^{\dagger AIl} Y_L^{AJn} Y_R^{\dagger AIn} Y_R^{AJl}
D_0(m_{u_A}^2, m_{u_A}^2, M_{H^+_l}^2, M_{H^+_n}^2)\phantom{aa}\nonumber\\
&&C_5(\mu) = 2\frac{(V_{AI}^\ast V_{AJ})^2}{16 \pi^2}\sum_A\sum_{l,n}^2
Y_L^{\dagger AIl} Y_L^{AJn} Y_R^{\dagger AIn} Y_R^{AJl}
D_{00}(m_{u_A}^2,m_{u_A}^2,M_{H^+_l}^2,M_{H^+_n}^2)
\end{eqnarray}
\normalsize The above results of the Wilson Coefficients are in
agreement with the ones in \cite{buras01}, the different sign and
factor are due to the different definitions of the four-point
integral functions. At one loop level there are no contributions
to the Wilson coefficients of the operators $Q_3$ and
$\tilde{Q}_3$.

In order to calculate the NLO QCD corrections to the Wilson
coefficients $C_i$ in the evolution from the scale of new physics
$\mu_t$ down to the low hadronic scale $\mu$ ($\sim m_b$), one has
to solve the corresponding renormalization group equations. The
details can be found in Ref.~\cite{renormalization}, with the
result
\begin{equation}
\label{eq:magic1}
C_r(m_b^{pole})=\sum_i \sum_s
  \left(b^{(r,s)}_i + \eta \,c^{(r,s)}_i\right) \eta^{a_i} \,C_s(\mu)
\end{equation}
where we have set the new physics scale $\mu_t = m_t$ and
$\eta=\alpha_s(\mu)/\alpha_s(m_t)$. The magic numbers $a_i$,
$b_i^{(r,s)}$ and $c_i^{(r,s)}$ are given in
Ref.~\cite{renormalization}.

\subsection{The Master Formulae}

In terms of the matrix elements of the effective $\Delta B =2$
Hamiltonian we have the neutral B meson mass difference
\begin{equation}
\Delta M_{B_q}=2|M_{12}^{(q)}|
\end{equation}
and the ratio of mixing parameters\footnote{For the sake of
simplicity we shall suppress the subscript $B_q$ of q/p
hereafter.}
\begin{equation}
(q/p)_{B_q}=-exp\left\{-i Arg [M_{12}^{(q)}]\right\} \label{qp}
\end{equation}
with
\begin{equation}
M_{12}^{(q)}=\langle \bar{B}^0_q | {\cal{H}}_{eff}(\Delta
B=2)|B^0_q \rangle.
\end{equation}

In the SM, the mass splitting $\Delta M_{B_q}$ at the NLO
level is available~\cite{buras96}
\begin{equation}
\Delta M_{B_q}  = \frac{G_F^2 M_W^2}{6 \pi^2}  m_{B_q}
(\hat{B}_{B_q} f_{B_q}^2 ) \eta_B S_0(x_t) |V_{tq}V_{tb}^\ast|^2
\end{equation}
where $S_0(x)$ is the Inami-Lim function~\cite{Inami-Lim}, the details of
$\hat{B}_{B_q}$ and $\eta_B$ can be found in Ref.~\cite{buras96}.

In the framework of model III 2HDM, we need to know the matrix
elements of the relevant operators $Q_i$ and $\tilde{Q}_i$ between
neutral B mesons. 
One defines~\cite{renormalization}
\begin{eqnarray}
\label{eq:bpars}
\langle \bar B_q \vert  Q_{1} (\mu)
\vert B_q \rangle & = &
\frac{1}{3} m_{B_q} f_{B_q}^{2}  B_1^{(q)}(\mu)
\nonumber \\
\langle
\bar B_q \vert Q_{2} (\mu) \vert B_q \rangle &=& -\frac{5}{24}
\left( \frac{ m_{B_q} }{ m_{b}(\mu) + m_q(\mu) }\right)^{2}
 m_{B_q} f_{B_q}^{2}  B_2^{(q)}(\mu) \nonumber  \\
\langle \bar B_q \vert Q_{3} (\mu) \vert B_q \rangle &=&
\frac{1}{24} \left( \frac{ m_{B_q} }{ m_{b}(\mu) + m_q(\mu) }\right)^{2}
 m_{B_q} f_{B_q}^{2}  B_{3}^{(q)}(\mu) \nonumber\\
\langle \bar B_q \vert Q_{4} (\mu) \vert B_q\rangle &=& \frac{1}{4}
\left( \frac{ m_{B_q} }{ m_{b}(\mu) + m_q(\mu) }\right)^{2}
 m_{B_q} f_{B_q}^{2} B_{4}^{(q)}(\mu) \nonumber\\
\langle \bar B_q \vert Q_{5} (\mu) \vert B_q \rangle &=&
\frac{1}{12} \left( \frac{m_{B_q}  }{ m_{b}(\mu) + m_q(\mu) }\right)^{2}
 m_{B_q} f_{B_q}^{2}  B_{5}^{(q)}(\mu)
\end{eqnarray}
where $f_{B_q}$ is the decay constant and $B_i$ is the so-called
bag factor. The matrix elements of $\tilde{Q}_{1-3}$ are the same
as that of $Q_{1-3}$. The lattice calculations of $B_i^{(q)}(\mu)$
have been done in Ref.\cite{bagpara} and the results are
\begin{equation}
\begin{tabular}{ll}
$B_{1}^{(d)}(m_b) =0.87(4)^{+5}_{-4}$ \;\;\;
    & $B_{1}^{(s)}(m_b) =0.86(2)^{+5}_{-4}$  \\
$B_{2}^{(d)}(m_b) = 0.82(3)(4)$ \;\;\;
    & $B_{2}^{(s)}(m_b) = 0.83(2)(4)$ \\
$B_{3}^{(d)}(m_b)=1.02(6)(9)$ \;\;\;
    & $B_{3}^{(s)}(m_b) = 1.03(4)(9)$ \\
$B_{4}^{(d)}(m_b) = 1.16(3)^{+5}_{-7}$ \;\;\;
    & $B_{4}^{(s)}(m_b) = 1.17(2)^{+5}_{-7}$ \\
$B_{5}^{(d)}(m_b) = 1.91(4)^{+22}_{-7}$ \;\;\;
    & $B_{5}^{(s)}(m_b) = 1.94(3)^{+23}_{-7}$
\end{tabular}
\end{equation}

\section{Numerical Analysis}

In numerical calculations, the following values of parameters are
assumed:
\begin{equation}
m_{B_d}=5.279 \, GeV, \;\; m_{B_s}=5.370 \, GeV, \;\;
m_{H^\pm}=200 \, GeV
\end{equation}
The values of $f_{B_q}$ and $\hat{B}_{B_q}$ in the lattice QCD
calculation and in QCD sum rules have been estimated in so many
works, see, e.g.~\cite{wittig,becirevic,battaglia,beneke}. Here,
we quote the results listed in \cite{wittig}.
\begin{eqnarray}
f_{B_d} &=& 191 \pm 23^{+0}_{-19} \, MeV
\nonumber\\
f_{B_s} &=& 220 \pm 25 \, MeV
\nonumber\\
f_{B_d} \sqrt{\hat{B}_{B_d}} &=& 221 \pm 28^{+0}_{-22} \, MeV
\nonumber\\
f_{B_s}\sqrt{\hat{B}_{B_s}} &=& 255 \pm 31 \, MeV
\end{eqnarray}

In \cite{hll} we calculated the constraint on $|\lambda_{bb}|$ and
$|\lambda_{tt}|$ due to the experimental upper bound of $Br (B_s \rightarrow
\mu^+\mu^-)$ which is~\cite{pdg2003}
\begin{equation}
B_r (B_s \rightarrow \mu^{+}\mu^-) < 2.0 \times 10^{-6} (CL=90\% )
\end{equation}
Recently, the Ref.~\cite{cdf04} gave the new bound of $Br (B_s \rightarrow
\mu^+\mu^-)$ which reads
\begin{equation}
B_r (B_s \rightarrow \mu^{+}\mu^-) < 5.8 \times 10^{-7} (CL=90\% )
\end{equation}
With the updating upper bound of $Br (B_s \rightarrow \mu^+\mu^-)$,
we recalculate the constraint on $|\lambda_{bb}|$ and $|\lambda_{tt}|$
and the result is shown in Fig.1.
The cyan region is allowed by the new experimental bound of $Br
(B_s \rightarrow \mu^+\mu^-)$. Comparing with the Fig.3 in
Ref.~\cite{hll}, we find that the new bound gives a more stringent
constraint on the values of $|\lambda_{bb}|$ and $|\lambda_{tt}|$,
as expected.

It is shown in Ref. \cite{bck} that the strictest constraints on
$\lambda_{bb}$ and $\lambda_{tt}$ come from $B \rightarrow
X_s\gamma$ and the neutron electric dipole moment (NEDM).
Considering the theoretical uncertainties, we take $2.0\times
10^{-4} < {\rm Br}(B\to X_s \gamma)< 4.5\times 10^{-4}$, as
generally analyzed in literatures. The NEDM can be expressed as
\begin{equation}
d^g_n=10^{-25}\hbox{e$\cdot$cm} \  \hbox{Im}(\lambda_{tt}
\lambda_{bb}) \left({\alpha(m_n)\over\alpha(\mu)}\right)^{1\over2}
\left({\xi_g\over 0.1}\right) H\left({m_t^2\over
M^2_{H^\pm}}\right)
\end{equation}
with
\begin{equation}
H(y)={3\over2}{y\over (1-y)^2} \left(y-3-{2\log y\over 1-y}\right)
\end{equation}
Here the parameter $\xi_g$ has been estimated and has different
values obtained by different methods: 0.03, 0.07 and 1, which is
due to the hadronic uncertainty~\cite{bck}. Using the result in
Ref.\cite{bck}, considering the uncertainty of $\xi_g$ and
finding the maximally possible $|\lambda_{bb}\lambda_{tt}|$ which
satisfies the experimental constraints, and requiring that the
charged Higgs boson is not too heavy (say, $m_{H^\pm}<250 GeV$) and the
phase of $\lambda_{bb}\lambda_{tt}$ is not too limited, it follows
that $|\lambda_{bb}\lambda_{tt}|\sim 3$.

From the well measured physical observable $\Delta M_{B_d}$ and
using the values of $|\lambda_{bb}|$ and $|\lambda_{tt}|$ analyzed
above, one can get the constraint on $|\lambda_{cc}|$ in the model
III. In Fig.2, we show the correlation between $|\lambda_{cc}|$
and $\theta_{bb}$ ($\theta_{ii}$ is the phase of $\lambda_{ii}$)
for $\theta_{bb}+\theta_{tt}=\pi/2$\footnote{The terms relevant to
$\theta_{cc}$ in the formula of $\Delta M_{B_d}$ have been
neglected because they are very small compared with other terms.}
due to the constraint from $\Delta M_{B_d}$ within the $1\sigma$
deviation (i.e., $\Delta M_{B_d}=0.502 \pm 0.007 ps^{-1}$). One
can see from the figure that the constraint on $|\lambda_{cc}|$ is
very stringent and the dependence of $|\lambda_{cc}|$ on
$\theta_{bb}$ is weak. We have also carried out the calculations
for different values of $\theta_{bb}+\theta_{tt}$ in the allowed
range and the results are similar, i.e., the dependence of
$|\lambda_{cc}|$ on both $\theta_{bb}$ and $\theta_{tt}$ is weak.

With the experimental lower bound of $\Delta M_{B_s}$, i.e.,
$\Delta M_{B_s}>14.4 ps^{-1}$, we can draw the lower bound on
$|\lambda_{ss}|$. Fig.3 is the contour plot of $|\lambda_{ss}|$
versus $\theta_{ss}$ for fixed $|\lambda_{tt}|$, $|\lambda_{bb}|$
and $|\lambda_{cc}|$ which satisfied the constraints from
$B_d-\bar{B}_d$ mixing , $\Gamma(b \to s\gamma)$, $\Gamma(b \to c
\tau \bar{\nu}_\tau)$, $\rho_0$, $R_b$ and the electric dipole
moment of neutron. The yellow area is excluded by the measured
experimental bound. Above the yellow region, all the values of
$|\lambda_{ss}|$ are allowed. We hope that the future B factory
can make precise measurements of $\Delta M_{B_s}$, consequently,
the more stringent constraints on the free parameters in the model
III can be obtained.

Using the values of parameters obtained by analyzing above, we
calculate the new contribution to the ratio $q/p$. The Figs. 4 and
5 are devoted to $Arg[(q/p)_n]$ which denotes the new contribution
to the phase of $q/p$ versus $\theta_{cc}$ for $B_s$ and $B_d$
systems respectively. As for the $B_s$ system, in Fig.4, we show
the $Arg[(q/p)_n]$ dependence on $\theta_{cc}$ for fixed
$|\lambda_{cc}|=100$ and for $|\lambda_{ss}|=80$ (solid curve),
100 (short-dashed curve) and 120 (dotted curve) respectively. From
the figure we find $Arg[(q/p)_n]$ increases with $|\lambda_{ss}|$
increasing. In particular, $Arg[(q/p)_n]$ is large enough to give
significant effects on CP violation in the neutral $B_s$ system.
In Fig.5, for the $B_d$ system, the solid curve corresponds
$|\lambda_{cc}|=98$ and the dotted one corresponds
$|\lambda_{cc}|=101$. We find the phase of $(q/p)_n$ is very
small, which verifies the expectation in Ref.\cite{hll}, and
consequently in agreement with the measurements of the time
dependent CP asymmetry $S_{J/\psi K}$ in $B\to J/\psi K_S$. In
contrast with the case of $B_s$, $Arg[(q/p)_n]$ decreases while
$|\lambda_{cc}|$ increasing.

\section{Conclusions}
In summary, we have calculated the new physics contributions to
the neutral B meson mass splitting $\Delta M_{B_q}$ (q=d, s) at
the NLO level in the model III 2HDM, while taking into account the
new CP violating complex phases involved Higgs bosons and fermions
coupling. By comparing the theoretical predictions with the high
accuracy data as well as other relevant data, we have drawn the
constraints on the free parameters of model III. Moreover, by
using the constrained parameters we calculated the corrections to
the ratio $q/p$. It is found that the phase of $(q/p)_n$ for $B_d$
which is due to the new contributions is very small and
consequently in agreement with the measurements of the time
dependent CP asymmetry $S_{J/\psi K}$ in $B\to J/\psi K_S$. On the
contrary, the phase of $(q/p)_n$ for $B_s$ is large enough to give
significant effects on CP violation in the neutral $B_s$ system.

\begin{acknowledgments}
One of the authors J.-T. Li would like to thank Dr. X.-H. Wu and J.-J. Cao
for enlightened discussions. This work was supported in part by the National
Nature Science Foundation of China.
\end{acknowledgments}

\newpage

\section*{\bf Appendix}

The Passarino-Veltman one-loop four-point functions with zero external
momenta are defined as
\begin{equation}
\int \frac{d^4q}{i
\pi^2}\frac{1}{(q^2-m_0^2)(q^2-m_1^2)(q^2-m_2^2)(q^2-m_3^2)}
=D_0(m_0^2,m_1^2,m_2^2,m_3^2)
\end{equation}

\begin{equation}
\int \frac{d^4q}{i \pi^2}\frac{q_\mu
q_\nu}{(q^2-m_0^2)(q^2-m_1^2)(q^2-m_2^2)(q^2-m_3^2)}
=g_{\mu \nu}D_{00}(m_0^2,m_1^2,m_2^2,m_3^2)
\end{equation}
where
\begin{equation}
D_0(m^2_0,m^2_1,m^2_3,m^2_3) = \frac{1}{6} \frac{1}{m^4_3}
f_{d0}(x,y,z) \hspace{12mm}x=\frac{m^2_0}{m^2_3},
y=\frac{m^2_1}{m^2_3},
z=\frac{m^2_2}{m^2_3}
\end{equation}
\begin{equation}
D_{00}(m^2_0,m^2_1,m^2_2,m^2_3) = -\frac{1}{12} \frac{1}{m^2_3}
f_{d00}(x,y,z) \hspace{10mm}x=\frac{m^2_0}{m^2_3},
y=\frac{m^2_1}{m^2_3},z=\frac{m^2_2}{m^2_3}
\end{equation}
with
\begin{eqnarray}
f_{d0}(x,y,z)&=& 6 \left[ \frac{x\ln x}{(1-x)(x-y)(x-z)} +
\frac{y\ln y}{(1-y)(y-x)(y-z)}\right. \ \nonumber\\
&+& \left. \frac{z\ln z}{(1-z)(z-x)(z-y)}\right]
\end{eqnarray}
\begin{eqnarray}
f_{d00}(x,y,z) &=& -3 \left[ \frac{x^2\ln x}{(1-x)(x-y)(x-z)} +
\frac{y^2\ln y}{(1-y)(y-x)(y-z)}\right. \ \nonumber\\
&+& \left. \frac{z^2\ln z}{(1-z)(z-x)(z-y)}\right]
\end{eqnarray}


\newpage

\vspace*{-1cm}


\begin{figure}[t]

\begin{center}

\epsfig{file=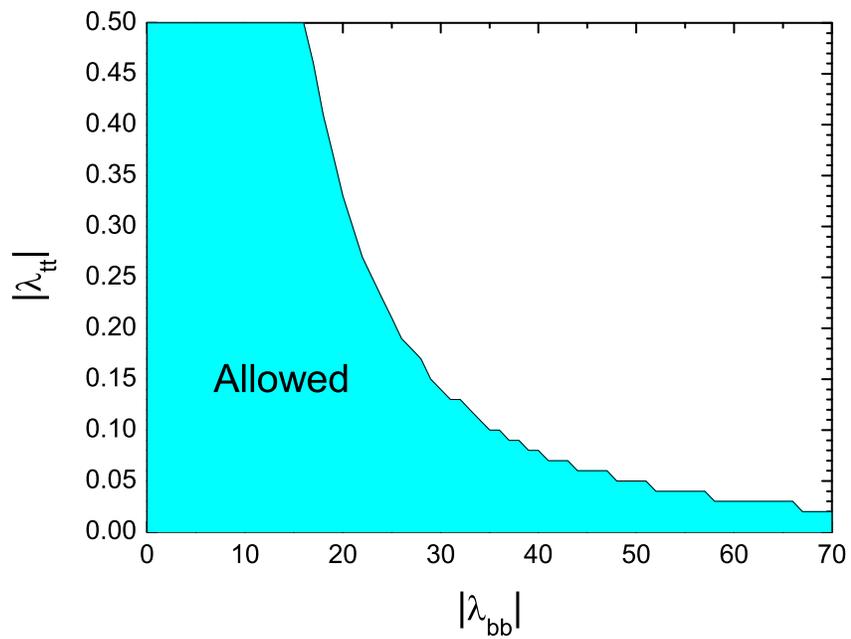,width=13cm} \vspace*{0.5cm}
\caption{The constraint on $|\lambda_{bb}|$ and $|\lambda_{tt}|$
due to the new experimental upper bound on $Br(B_s\rightarrow
\mu^+\mu^-)$ which reads $Br(B_s\rightarrow \mu^+\mu^-) < 5.8
\times 10^{-7} (CL=90\%)$.}

\label{fig: boundupdate}
\end{center}

\end{figure}


\vspace*{-1cm}


\begin{figure}[b]

\begin{center}

\epsfig{file=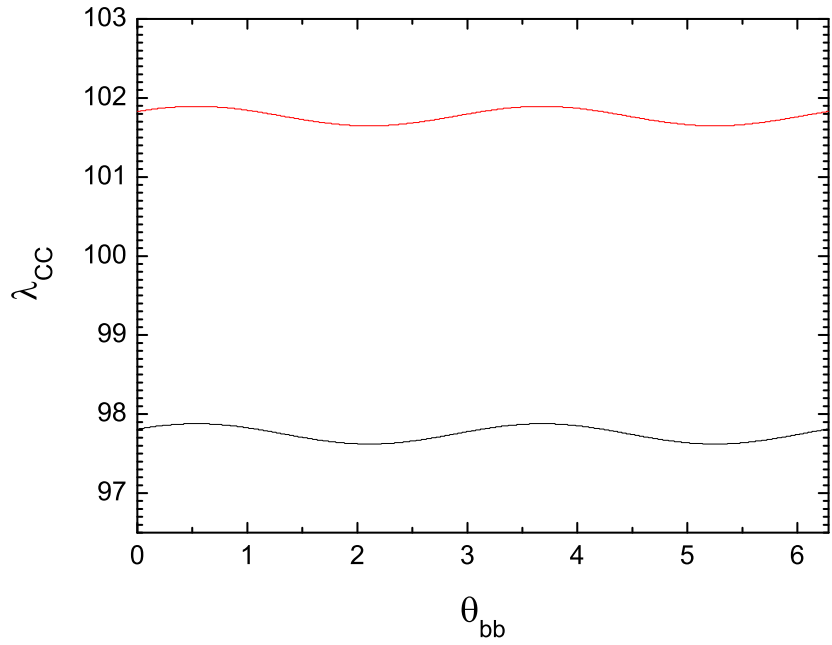,width=13cm} \vspace*{0.5cm} \caption{The
constraint on $|\lambda_{cc}|$ due to the measured $\Delta
M_{B_d}$ within $1\sigma$ deviation. }

 \label{fig:lambdacc}

\end{center}

\end{figure}


\vspace*{-1cm}


\begin{figure}[b]

\begin{center}

\epsfig{file=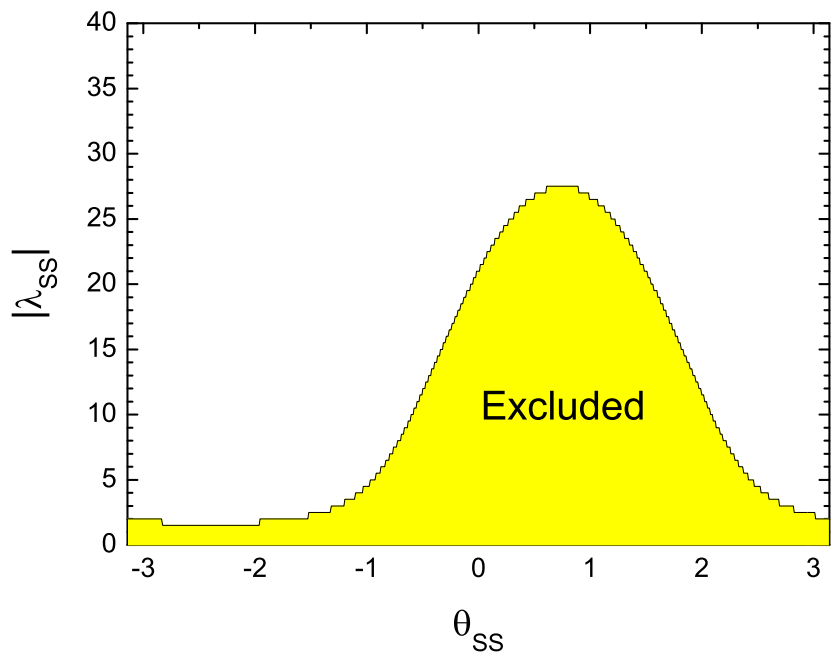,width=13cm} \vspace*{0.5cm}

\caption{The constraint on $|\lambda_{ss}|$ due to the lower experimental
bound on $\Delta M_{B_s}$.}

\label{fig:lambdass}

\end{center}

\end{figure}


\vspace*{-1cm}


\begin{figure}[b]

\begin{center}

\epsfig{file=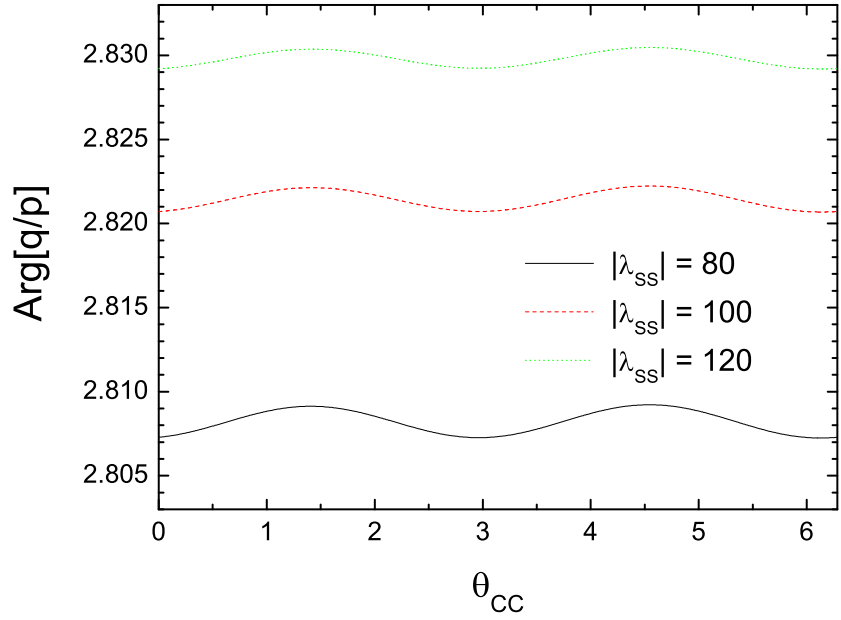,width=13cm} \vspace*{0.5cm}

\caption{$Arg[(q/p)_n]$ for $B_s$ system versus the CP violating
phase $\theta_{cc}$, for $|\lambda_{cc}|=100$. The solid curve
stands for $|\lambda_{ss}|=80$, the short-dashed curve stands for
$|\lambda_{ss}|=100$ and the dotted curve stands for
$|\lambda_{ss}|=120$ respectively. }

\label{fig:argqpbs}

\end{center}

\end{figure}


\vspace*{-1cm}


\begin{figure}[b]

\begin{center}

\epsfig{file=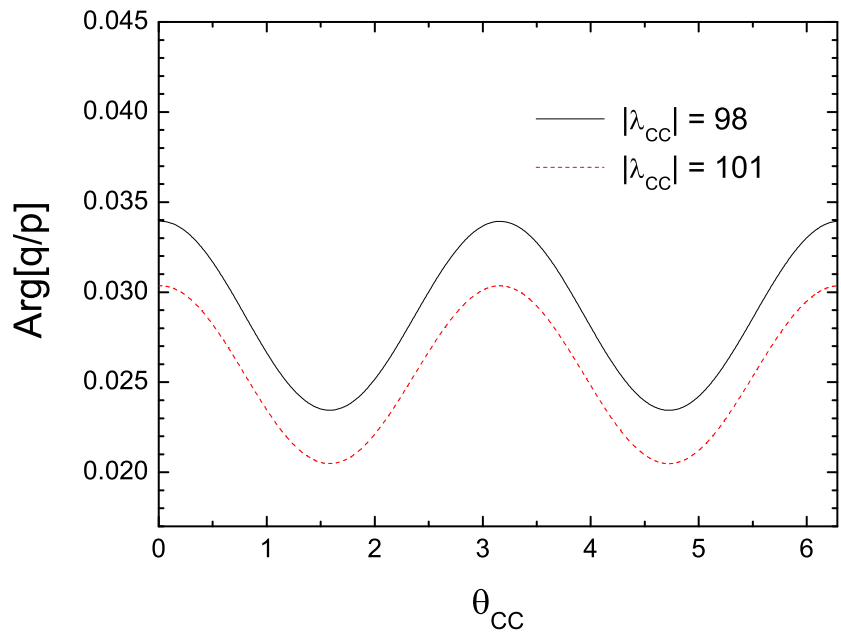,width=13cm} \vspace*{0.5cm}
\caption{$Arg[(q/p)_n]$ for $B_d$ system versus CP violating phase
$\theta_{cc}$ for $|\lambda_{cc}|=98$ (the solid curve) and for
$|\lambda_{cc}|=101$ (the dotted curve) respectively. }

\label{fig:argqpbd}

\end{center}
\end{figure}

\end{document}